\documentclass[prb,twocolumn,showpacs,amsmath,amssymb]{revtex4-1}

\usepackage[dvipdfmx]{graphicx}
\usepackage{bm}
\usepackage{color}

\usepackage{braket}

\begin{document}


\title{Generalized bulk-edge correspondence for non-hermitian topological systems}

\author{Ken-Ichiro Imura}
\author{Yositake Takane}

\affiliation{Department of Quantum Matter, AdSM, Hiroshima University, 739-8530, Japan}

\date{\today}


\begin{abstract}
A modified periodic boundary condition adequate for non-hermitian 
topological systems is proposed.
Under this boundary condition a topological number characterizing the system 
is defined in the same way as in the corresponding hermitian system
and hence,
at the cost of introducing an additional parameter that characterizes the non-hermitian skin effect,
the idea of bulk-edge correspondence in the hermitian limit
can be applied almost as it is.
We develop this framework through the analysis of
a non-hermitian SSH model with chiral symmetry,
and prove the bulk-edge correspondence
in a generalized parameter space. 
A finite region in this parameter space with a nontrivial pair of chiral winding numbers
is identified as topologically nontrivial,
indicating the existence of a topologically protected edge state under open boundary. 
\end{abstract}


\maketitle

\section{Introduction}
Non-hermiticity in quantum mechanics has been discussed since some time.
\cite{feshbach,NH1,NH2,bender}
A new trend in this field is its conjunction with the field of topological insulator.
\cite{GF_NH,ozawa,PRX,
Msato,hughes,
tony,YaoWang,non-Bloch,
anom_loc,biortho,why,svd,transfer_M,
okuma,
hide_qwalk2,
simon,KK_chern,
Fu,nori,yuce,bbc_1d,bipolar,gil,thomale,
doubled,hybrid,ring,AAH,
KK38,KK_unif,
hide_natphys,
SSH_GL_laser,SSH_GL_nmat,TI_laser1,TI_laser2,
silicon,nphys_QW,photo,nature,science,
RLC1,RLC2}
An intensive research on non-hermitian topological systems
includes quite a few experimental studies, 
\cite{hide_natphys,
SSH_GL_laser,SSH_GL_nmat,TI_laser1,TI_laser2,
silicon,nphys_QW,photo,nature,science,
RLC1,RLC2}
implying that
non-hermitian topological physics may not only be interesting but also useful; 
{\it e.g.}, topological insulator laser.
\cite{SSH_GL_laser,SSH_GL_nmat,TI_laser1,TI_laser2,
silicon,nphys_QW,photo,nature,science}
Non-hermiticity requires a completely new point of view
exotic to the hermitian world, 
{\it e.g.}, the biorthogonal approach, \cite{brody,biortho,Fu}
stimulating theoretical studies in different directions.
\cite{NH3,NH4,heiss,osaka1,KK_ex,
osaka2,
landau,papaj,kozii,
tsuneya,tsuneya_X,
magnon}

The non-hermitian system
is sensitive to boundary conditions.\cite{NH2,PRX}
So is the topological system, but in a totally different way.
In topological systems,
one can make the edge states appear or disappear
by controlling the boundary condition.
With the relevant topological number defined and
evaluated
in the periodic boundary condition
at hand,
whether the edges states appear or not in the system of open boundary
is predestined (bulk-edge correspondence).\cite{HG,ryu}
Here, in the non-hermitian systems we consider,
the change of the boundary condition has
a more profound impact
even on the bulk physics;
not only the edge but also
the bulk part of the spectrum 
and the corresponding wave functions
are susceptible to a qualitative change
in the application of a different boundary condition.

\begin{figure}
\includegraphics[width=85mm, bb=0 0 252 107]{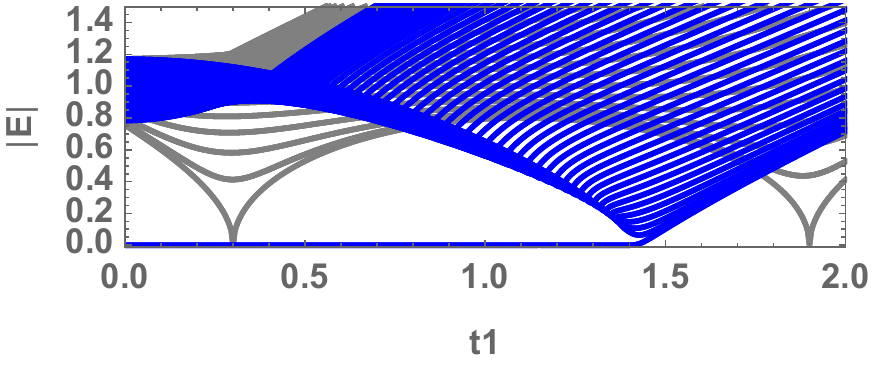}
\vspace{-2mm}
\caption{
A typical energy spectrum of the non-hermitian SSH model [see Eq. (\ref{H_obc})]
under open boundary. 
The absolute value $|E|$ of the complex energy spectrum is
shown in blue with varying $t_1$. 
Other parameters are 
$t_2 = 1$, $t_3 = 1/10$, $\gamma_1 = 3/4$, $\gamma_2 = 1/20$.
The spectrum is obtained by numerically diagonalizing $H_{obc}$ at $2L=100$.
For comparison, the spectrum of the same model
under periodic boundary is shown in gray.
}
\label{spec_obc}
\end{figure}

Terminologies such as bulk and edge geometries are 
convenient in considering topological systems.
In hermitian systems,
applying
an open (periodic) boundary condition 
is equivalent to
placing it in the edge (bulk) geometry.
A quantum system is called topological when
it exhibits protected gapless (or zero-energy) edge states
in the edge geometry $g_{edge}$,
while the corresponding topological number 
is defined in the bulk geometry $g_{bulk}$.
Bulk-edge correspondence assures a one-to-one relation
between
the system's behavior under
$g_{edge}$ and under $g_{bulk}$.
In non-hermitian systems
this fundamental principle becomes elusive, at least superficially.
\cite{hughes,Msato,biortho,why,svd,transfer_M,PRX}
Examples are known in which
the number and locations
of gap closing (topological phase transition) 
differ under open and periodic boundary conditions
(see Fig.~\ref{spec_obc}),
i.e.,
the system's behavior under $g_{edge}$ 
is no longer constrained by that under $g_{bulk}$.
Such a feature is typical to
non-hermitian system with anisotropic hopping.
The underlying reason is the non-hermitian skin effect
which refers to the fact that
under open boundary condition
even the bulk wave function tends to be localized
in the vicinity of an edge.
\cite{tony,YaoWang,biortho}
The key to recover the bulk-edge correspondence is to reconsider the setting 
of $g_{bulk}$;
the boundary condition concretizing $g_{bulk}$ needs to be readjusted
in the presence of non-hermitian skin effect.

The purpose of this paper is to promote the use of the boundary condition
which we dub
modified periodic boundary condition (mpbc).
The mpbc is the most appropriate boundary condition for realizing $g_{bulk}$
in non-hermitian topological systems 
that exhibit non-hermitian skin effect.
Our scenario is illustrated by applying mpbc to the analysis of 
a non-hermitian SSH model
with anisotropic hopping.
\cite{YaoWang,non-Bloch}
We apply the proof of the bulk-edge correspondence of Ref. \onlinecite{ryu}
to this model
with the use of mpbc.
Under mpbc
a pair of topological winding numbers $w_\pm$ [see Eq. (\ref{wn_pm})]
are introduced to characterize the system.
Our mpbc contains a free parameter $b$.
Therefore, 
$w_\pm$ are the functions 
not only of the original set of parameters $\tau=\{t_\mu,\gamma_\mu\}$ ($\mu=1,2,3,\cdots$)
[see, e.g., Eqs. (\ref{H_obc}),(\ref{tau12})],
but also of $b$.
By evaluating $w_\pm$,
a finite region in the generalized parameter space $\tilde{\tau}=\{\tau,b\}$
is identified to be topologically nontrivial,
which is also smoothly connected to the known topological insulator (TI) region in the hermitian limit.
In the same phase diagram 
we have also identified 
in addition to
ordinary insulator (OI) regions,
new topological regions 
specific to non-hermitian systems
that have no analog in the hermitian limit.

\section{The modified periodic boundary condition}

In a non-hermitian system even the 
bulk wave function 
does not necessarily extend over the entire system.
To illustrate this 
let us consider 
the following 1D tight-binding model with an anisotropic hopping $t\neq t'$; the so-called Hatano-Nelson model \cite{NH1,NH2}
in the clean limit:
\begin{equation}
H_1^{obc} =\sum_{j=1}^L (t |j+1\rangle\langle j| + t' |j\rangle\langle j+1|),
\label{obc}
\end{equation}
where
$|j\rangle$
represents a state localized at the $j$th site.
In Eq. (\ref{obc})
an open boundary condition (obc)
is implicit, since
hopping amplitude are truncated at $j=1$ and $j=L$.
Under obc,
the wave function of the system
tends to be localized as
\begin{equation}
\psi_j \propto b^j \sin kj,
\label{NHSE}
\end{equation}
where $b=\sqrt{t/t'}$, $k=n\pi/(L+1)$ with $n=1,2,\cdots,L$.
Compared with the hermitian case: $t=t'$,
the wave function (\ref{NHSE}) is multiplied by a factor $b^j$.
Such a behavior, referred to as non-hermitian skin effect, 
manifests under obc.
The quantity $b$
measures the degree of amplification (attenuation) of the wave function
due to non-hermitian skin effect.


To establish a bulk-edge correspondence
the existence of a reference surfaceless geometry is mandatory.\cite{ryu}
A winding number insuring 
the emergence of a zero-energy edge state under 
obc is defined in this geometry.
In hermitian systems this role is 
concretized by the periodic boundary condition (pbc).
In non-hermitian systems 
the ordinary pbc
sometimes fails to play the same role.
\cite{YaoWang,biortho}
The positions of gap closing
is an important part of
the topological information on the system.
Fig. 1 shows that in this non-hermitian system
such an information 
does not seem to be kept properly
when the boundary condition is changed from open to periodic.

The difficulty 
of the ordinary pbc
stems from the fact that
it fails to take proper account of the non-hermitian skin effect.
To overcome this difficulty, we propose to use the following boundary Hamiltonian:
\begin{equation}
\Delta H_1^{mpbc} = b^{-L} t |1\rangle\langle L| + b^L t |L\rangle\langle 1|,
\label{mpbc1}
\end{equation}
which represents a modified periodic boundary condition (mpbc) 
[as for the naming see text after Eq. (\ref{mpbc2})].
The matrix elements of Eq. (\ref{mpbc1})
are to be added to the obc Hamiltonian (\ref{obc}).
From an infinite set of fundamental solutions, 
\begin{equation}
|\beta\rangle = \sum_j \beta^j |j\rangle,
\label{beta1}
\end{equation}
of the infinite system,
the factor $b^{\pm L}$ filters out
those satisfying $|\beta|=b$, which 
mimic in a controllable manner
the spatial amplification or attenuation 
of the wave function (\ref{NHSE})
under obc.
Note that 
Eq. (\ref{mpbc1}) is non-hermitian unless $b=1$.

The 
mpbc can be formulated
on a more generic basis.
To impose mpbc is equivalent to require
\begin{equation}
\psi (j+L) = b^L \psi (j),
\label{mpbc2}
\end{equation}
to wave functions in an infinite system, 
in the sense that the resulting wave functions
in any interval of $L$ sites are equivalent to those in the system of $L$ sites with mpbc.
Note that such wave functions in the infinite system are not bounded
and hence are 
not allowed in quantum mechanics, 
whereas those in the finite system have no such
difficulty.
Let $\psi_j$ be a wave functions that is selected by the condition (\ref{mpbc1}).
If one expresses this wave function as 
$\psi_j = b^j \phi_j$,
then the {\it rescaled} wave function $\phi_j$ 
satisfies the ordinary periodic boundary condition:
$\phi_{j+L} = \phi_j$,
and in this sense the present boundary condition (\ref{mpbc1}) 
may be called a modified periodic boundary condition.

To specify the mpbc (\ref{mpbc1}),
one has to specify the parameter $b$.
In the simple model prescribed by Eq. (\ref{obc}), 
$b$ is uniquely determined as $b = \sqrt{t/t'}$ to mimic the non-hermitian
skin effect. 
However, in a more generic model, such a plausible
value of $b$ generally depends on $k$ as each wave function
exhibits its own spatial amplification or attenuation under obc.
Taking these into account, below we consider $b$ as an independent
parameter. 
This in turn signifies that the space of
model parameters specifying our system has been slightly
enlarged: $\tau\rightarrow\tilde{\tau} = \{\tau, b \}$.
Later, we discuss the bulk-edge correspondence of our system in this generalized parameter space $\tilde{\tau}$.



\section{The model system}

To illustrate our scenario
we employ the SSH-type non-hermitian (tight-binding) Hamiltonians 
\cite{tony,YaoWang,simon}
The nearest-neighbor (NN) SSH model with anisotropic hopping
employed in Refs. \onlinecite{tony,YaoWang}
gives a prototypical example
in which one encounters the difficulty of applying the periodic boundary condition (pbc)
to a non-hermitian system.
Here, we consider a slightly generalized version of this model
with third-nearest-neighbor (3NN) hopping:
\cite{non-Bloch,YaoWang}
\begin{eqnarray}
H_{obc}&=&H_{\rm NN}+H_{\rm 3NN},
\label{H_obc}
\\
H_{\rm NN} 
&=& \sum_{j=1}^{L} 
\left[t_1^- |jB\rangle\langle jA| + t_1^+ |jA\rangle\langle jB|
\right]
\nonumber \\
&+&
\sum_{j=1}^{L-1}
\left[
t_2^-
|j+1,A\rangle\langle j,B| + 
t_2^+
|jB\rangle\langle j+1,A|\right],
\nonumber \\
H_{\rm 3NN} 
&=& \sum_{j=1}^{L-1} 
\left[t_3 |j+1,B\rangle\langle jA| + t_3 |jA\rangle \langle j+1,B|
\right],
\nonumber
\end{eqnarray}
where
\begin{equation}
t_1^\pm=t_1\pm\gamma_1,\ \
t_2^\pm=t_2\pm\gamma_2,
\label{tau12}
\end{equation}
represent intra- and inter-cell
anisotropic hopping amplitudes.
\footnote{Here, we inherit the convention of Ref. \onlinecite{YaoWang} and denote the hopping amplitude in the {\it forward} direction as $t_1^-$ and $t_2^-$.}
The total Hamiltonian $H_{obc}$ becomes hermitian in the limit: $\gamma_1=\gamma_2=0$.
Note that
the model parameters
$t_\mu (\mu=1,2,3)$,
$\gamma_\mu (\mu=1,2)$
are all real constants.

The open boundary condition (obc) is implicit in Eqs. (\ref{H_obc}),
since hopping amplitudes are truncated at $j=1$ and $j=L$.
The size of the system is $L$ in unit cells,
and $2L$ in the number of sites.
A typical energy spectrum of 
our model under open boundary is shown in Fig. 1
with varying $t_1$,
while other parameters are
$t_2 = 1$,
$t_3 = 1/10$,
$\gamma_1 = 3/4$,
$\gamma_2 = 1/20$.
The spectrum is obtained by numerically diagonalizing $H_{obc}$
at $2L=100$.

Now that our model is
explicitly given, 
we can also write down
the modified periodic boundary condition (mpbc) explicitly:
\begin{eqnarray}
\Delta H_{\rm NN}^{mpbc}
&=& 
b^{-L} t_2^-
|1,A\rangle\langle L,B| + 
b^L t_2^+
|L,B\rangle\langle 1,A|,
\nonumber \\
\Delta H_{\rm 3NN}^{mpbc}
&=& 
b^{-L} t_3 |1,B\rangle\langle L,A| + 
b^L t_3 |L,A\rangle \langle 1,B|.
\label{H_mpbc}
\end{eqnarray}
These boundary Hamiltonians are combined with the obc Hamiltonian $H_{obc}$
to give the mpbc Hamiltonian,
\begin{equation}
H_{mpbc}=H_{obc}+\Delta H_{\rm NN}^{mpbc}+\Delta H_{\rm 3NN}^{mpbc}.
\end{equation}
The hopping matrix elements in
Eqs. (\ref{H_mpbc})
connect the final unit cell $j=L$ back to the first one $j=1$,
and vice versa.
Note that
the boundary Hamiltonians (\ref{H_mpbc})
are non-hermitian unless $b=1$.

In the following transformed basis,\cite{NH1,NH2,YaoWang}
the 
mpbc (\ref{H_mpbc})
reduces to an ordinary pbc. 
Let $S$ be the transforming matrix:
\begin{equation}
S={\rm diag} [1,1,b,b,b^2,b^2,\cdots,b^{L-1},b^{L-1}],
\label{similar1}
\end{equation}
and consider the similarity transformation:
\begin{eqnarray}
\tilde{H}_{mpbc}&=&S^{-1}H_{mpbc}S
\label{similar2}
\\
&=&\tilde{H}_{obc}
+\Delta \tilde{H}_{\rm NN}^{mpbc}+\Delta \tilde{H}_{\rm 3NN}^{mpbc},
\nonumber 
\end{eqnarray}
where
$\tilde{H}_{obc}=\tilde{H}_{\rm NN}+\tilde{H}_{\rm 3NN}$
with
\begin{eqnarray}
\tilde{H}_{\rm NN} 
&=& \sum_{j=1}^{L} 
\left[t_1^- |jB\rangle\langle jA| + t_1^+ |jA\rangle\langle jB|
\right]
\nonumber \\
&+&
\sum_{j=1}^{L-1}
\left[
b^{-1}t_2^-
|j+1,A\rangle\langle j,B| + 
b t_2^+
|jB\rangle\langle j+1,A|\right],
\nonumber \\
\tilde{H}_{\rm 3NN} 
&=& \sum_{j=1}^{L-1} 
\left[b^{-1}t_3 |j+1,B\rangle\langle jA| + b t_3 |jA\rangle \langle j+1,B|
\right],
\nonumber
\end{eqnarray}
and
\begin{eqnarray}
\Delta \tilde{H}_{\rm NN}^{mpbc}
&=& 
b^{-1} t_2^-
|1,A\rangle\langle L,B| + 
b t_2^+
|L,B\rangle\langle 1,A|,
\nonumber \\
\Delta \tilde{H}_{\rm 3NN}^{mpbc}
&=& 
b^{-1} t_3 |1,B\rangle\langle L,A| + b t_3 |L,A\rangle \langle 1,B|.
\label{tilde_pbc1}
\end{eqnarray}
Note that
$\Delta \tilde{H}_{\rm NN}^{mpbc}+\Delta \tilde{H}_{\rm 3NN}^{mpbc}$
represents the ordinary pbc 
for $\tilde{H}_{obc}$.
In Sec. II we have seen that under (\ref{mpbc1})
the rescaled wave function $\phi_j$
satisfies the ordinary pbc.
Here, by a similarity transformation (\ref{similar2})
the boundary Hamiltonians (\ref{H_mpbc}) representing the mpbc 
are reduced to the ones representing the ordinary pbc.
These give us a good reason to call
our boundary condition (\ref{H_mpbc})
modified pbc.
Note that the similarity transformation (\ref{similar2})
keeps the eigenvalues unchanged.

\begin{figure}
(a)
\includegraphics[width=65mm, bb=0 0 250 252]{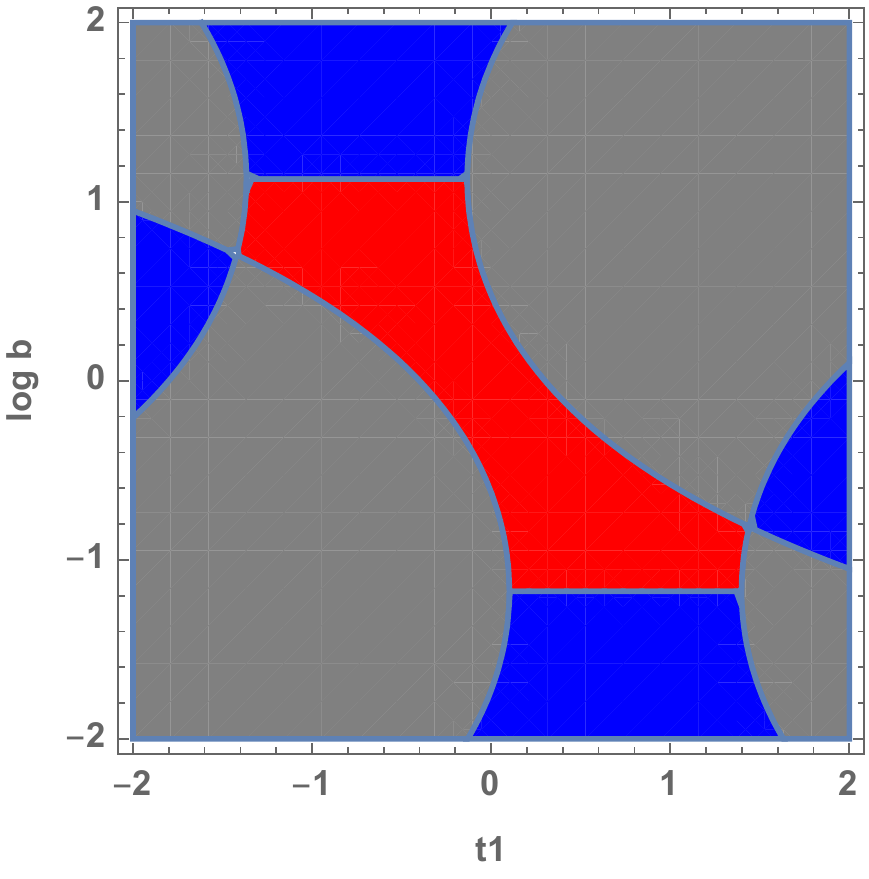}
\vspace{1mm}
\\
(b)
\includegraphics[width=65mm, bb=0 0 250 252]{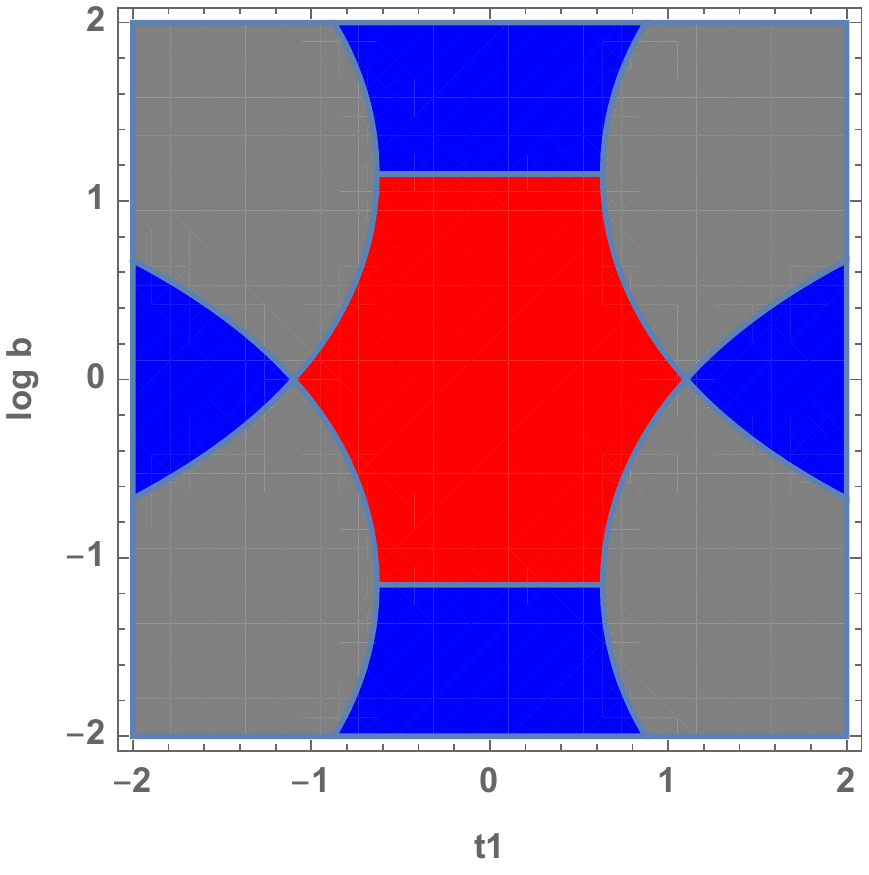}
\vspace{-2mm}
\caption{
The phase diagrams in the space of parameters $(t_1,b)$
determined by
the winding number $w(t_1,b)$
for (a) the non-hermitian case of 
$\gamma_1 = 3/4$,
$\gamma_2 = 1/20$
and (b) the hermitian case of $\gamma_1 = \gamma_2 = 0$.
Other parameters are
$t_2 = 1$,
$t_3 = 1/10$.
The regions of $w = 1$ and $w = 0$ are
respectively painted in red and blue. 
The remaining 
gray area corresponds
to $w = 1/2$.
}
\label{wn_map}
\end{figure}

\section{Generalized bulk-edge correspondence}

For a hermitian system with chiral (sublattice \cite{KK38}) symmetry
a clear proof of the bulk-edge correspondence is available.\cite{ryu}
In this proof
the Hamiltonian $H(g_{bulk})$ in a reference bulk
(closed, surfaceless) geometry $g_{bulk}$
is specified by a set of parameters, which correspond to $R_\pm$
in the following formulation.
The role of $g_{bulk}$ is played by the
ordinary periodic boundary condition (pbc)
so that
$H(g_{bulk})$ becomes the standard Bloch Hamiltonian $H(k)$,
which is specified by
$R_+(k)=R_-^*(k)=R(k)$.
As $k$ sweeps the entire Brillouin zone,
the trajectory of $R(k)$ forms a loop 
on this complex parameter plane.
The topological property of $H(k)$
is encoded in the winding property of this loop
with respect to the origin (a reference point).
To judge whether
the system is topologically trivial or not,
one attempts to deform this loop continuously into
a special one (e.g., to the one corresponding to $t_1=0$, $t_2=1$ and $t_3=0$),
at which
the existence of a zero-energy edge state is apparent
in the edge geometry $g_{edge}$,
i.e., under the open boundary condition (obc).
The point is that 
to establish a bulk-edge correspondence,
in addition to 
$g_{edge}$,
the existence of 
$g_{bulk}$ 
is mandatory,
and
$g_{edge}$ is realized by
a simple truncation of $g_{bulk}$ in real space.
In non-hermitian systems
ordinary pbc
is no longer qualified for such a reference geometry,
since it is incapable of capturing the nature of wave function 
that tends to amplify exponentially under obc.
Introduction of ordinary pbc
changes the nature of the
wave function so abruptly
that it even changes the topological nature of the system
as indicated in Fig. 1.

Here, we show that
the modified periodic boundary condition (mpbc)
introduced in the last section
provides with such a reference geometry
for a generic non-hermitian topological system.
%
Below, we show that
the proof of bulk-edge correspondence 
as given in Ref. \onlinecite{ryu}
can be safely applied to
a generic non-hermitian topological system, 
using the reference geometry specified by mpbc.

In parallel with the single-band case (see Sec. II)
the eigenstate of $H_{mpbc}$
satisfying
\begin{equation}
H_{mpbc} |\beta\rangle =E(\beta) |\beta\rangle,
\end{equation}
takes the following form:
\begin{eqnarray}
|\beta\rangle=\sum_{j=1}^{L} \beta^j (c_A |jA\rangle
+c_B|jB\rangle),
\label{beta>}
\end{eqnarray}
where
\begin{equation}
\beta=b e^{ik},
\label{beta}
\end{equation}
with $k$ being real and in the range of the Brillouin zone: 
$k\in [0,2\pi]$. 
The coefficients $c_A$ and $c_B$ in Eq. (\ref{beta>}) 
are
determined by the eigenvalue equation:
\begin{eqnarray}
H_{mpbc}(\beta)
\left[\begin{array}{c}
c_A\\c_B
\end{array}\right]
=E(\beta)
\left[\begin{array}{c}
c_A\\c_B
\end{array}\right],
\label{eve}
\end{eqnarray}
where
\begin{equation}
H_{mpbc}(\beta)=
\left[
 \begin{array}{cc}
0&R_+(\beta)\\
 R_-(\beta) &0
  \end{array}
\right]
\end{equation}
is our reference bulk Hamiltonian, 
which is also explicitly chiral (sublattice \cite{KK38}) symmetric,
and
\begin{eqnarray}
R_+(\beta) &=& t_1^+ + t_2^- \beta^{-1} + t_3 \beta,
\nonumber \\
R_-(\beta) &=& t_1^- + t_2^+\beta + t_3 \beta^{-1}.
\label{R_pm}
\end{eqnarray}
Then,
following  Ref. \onlinecite{ryu}, let us introduce
the winding numbers:
\begin{eqnarray}
w_\pm &=& {1\over 2\pi}[ \arg R_\pm (\beta) ]_{k=0}^{2\pi}
\nonumber \\
&=&{1\over 2\pi}[\phi_\pm (2\pi)-\phi_\pm (0)],
\label{wn_pm}
\end{eqnarray}
where
\begin{equation}
\phi_\pm (k) = {\rm Im} \log R_\pm (\beta),
\end{equation}
with the choice of the branch of $\log$ such that
$\phi_\pm (k)$ is continuous for 
$k\in [0,2\pi]$.
As $k$ sweeps the entire Brillouin zone 
at a fixed $b$,
the winding number $w_\pm$ measures
how many times 
the trajectory of $\rho=R_\pm (\beta=b e^{ik})$ encircles the origin 
in the complex $\rho$-plane
in the anti-clockwise direction.
The winding number such as the ones in Eq. (\ref{wn_pm})
is widely used for characterizing 
a hermitian model of class AIII, 
e.g., the SSH model.
\cite{book}

Fig. \ref{wn_map} (a) shows the phase diagram
in the parameter space $(t_1,b)$ 
determined by the winding numbers, $(w_+, w_-)$.
The parameter space $(t_1,b)$ 
represents a subspace of $\tilde{\tau}=\{\tau,b\}$
at which
other model parameters are fixed to the following values:
$t_2 = 1$,
$t_3 = 1/10$,
$\gamma_1 = 3/4$,
$\gamma_2 = 1/20$.
Here,
to make the phase diagram look simpler,
we employ the symmetric version of $w_\pm$
defined as
\begin{equation}
w = -{w_+ - w_- \over 2}.
\label{sym}
\end{equation}
In the hermitian limit
the hermiticity requires $R_- = R_+^*$,
so that $w_-$ is fixed to $- w_+$.
Therefore, $w=-w_+=w_-$;
i.e., the symmetrization is not indispensable. 

In Fig. \ref{wn_map} (a) 
painted in red
is the region of $w=1$, corresponding to
$(w_+, w_-)=(1,-1)$.
Painted in blue
is the region of $w=0$, corresponding to
$(w_+, w_-)=(0,0)$.
The remaining 
gray area corresponds to $w=1/2$,
representing the regions of either
$(w_+, w_-)=(1,0)$ or $(0,-1)$.
In non-hermitian systems,
$w_+$ and $w_-$ 
vary independently,
taking in principle any combination of integral values.
As a result,
the symmetric winding number (\ref{sym}) 
can take half-integral values.
Fig. \ref{wn_map} (b) shows a similar phase diagram
in the limit: $\gamma_1=\gamma_2=0$.
The regions of $w=1/2$ remains to exist 
even in this limit
provided $b\neq 1$,
while
they disappear at $b=1$.

\begin{figure}
(a)
\includegraphics[width=80mm, bb=0 0 250 109]{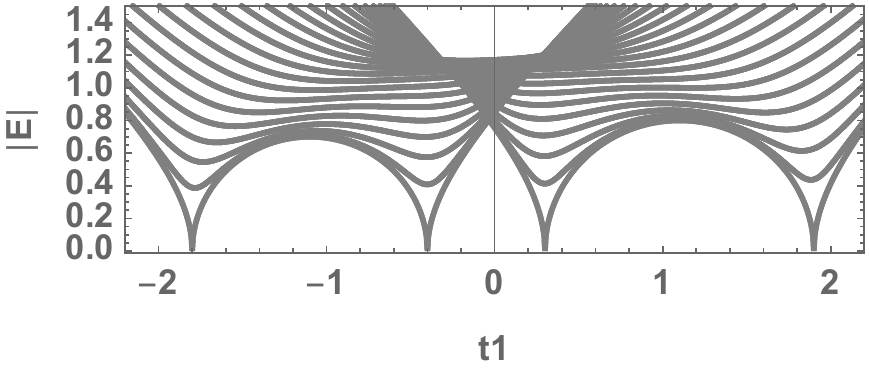}
\\
(b)
\includegraphics[width=80mm, bb=0 0 250 83]{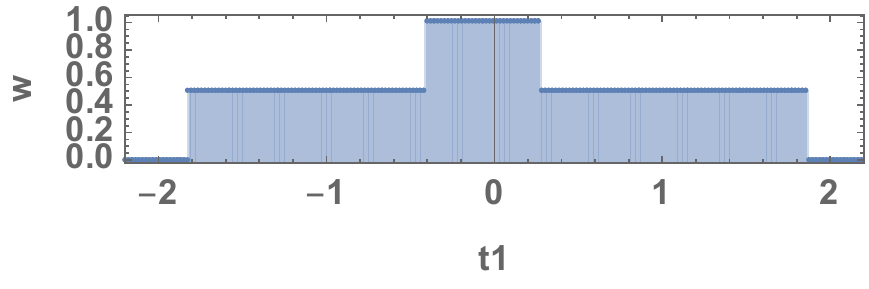}
\\
(c)
\includegraphics[width=80mm, bb=0 0 250 83]{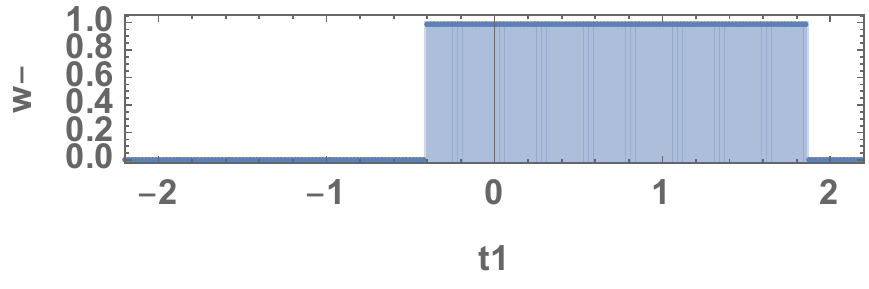}
\\
(d)
\includegraphics[width=80mm, bb=0 0 250 83]{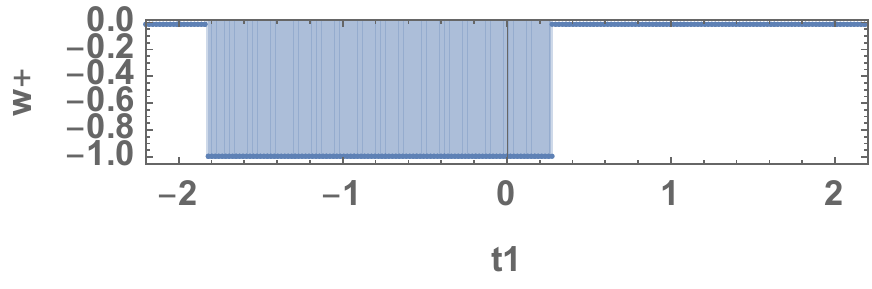}
\vspace{-2mm}
\caption{
The spectrum and winding numbers under pbc. 
In (a), $|E|$ is plotted as
a function of $t_1$. In (b)-(d), 
$w(t_1,b=1)$ and $w_\pm (t_1,b=1)$
are shown in
the same parameter range. 
Parameters other than $t_1$ and $b$ are same as those
in Fig. \ref{spec_obc}.
}
\label{spec_pbc}
\end{figure}


Let us argue that the two phase diagrams,
the one at a generic value of
$\gamma_1$ and $\gamma_2$
represented by Fig. \ref{wn_map} (a)
and
the one at $\gamma_1=\gamma_2=0$,
are smoothly connected.
On the phase boundaries of
different topological phases at a generic value of
$\gamma_1$ and $\gamma_2$ [Fig. \ref{wn_map} (a)]
either of $\rho=R_\pm (\beta)$ touches the origin 
in the complex $\rho$-plane; i.e., either of
\begin{eqnarray}
0&=&|R_-(\beta)|= |t_1-\gamma_1+ (t_2+\gamma_2)\beta + t_3 \beta^{-1}|,
\nonumber \\
0&=&|R_+(\beta)|=|t_1+\gamma_1 + (t_2-\gamma_2)\beta^{-1} + t_3 \beta|,
\label{rpm=0}
\end{eqnarray}
holds.
As the non-hermitian parameters $\gamma_1,\gamma_2$ vary,
these phase boundaries deform smoothly, and in the limit of
$\gamma_1=\gamma_2=0$,
Eqs. (\ref{rpm=0}) reduce to
\begin{eqnarray}
0&=&|R_-(\beta)|= |t_1+ t_2\beta + t_3 \beta^{-1}|,
\nonumber \\
0&=&|R_+(\beta)|=|t_1+ t_2\beta^{-1} + t_3 \beta|.
\end{eqnarray}
These conditions indeed define
the phase boundaries of
different topological phases at $\gamma_1=\gamma_2=0$ 
[Fig. \ref{wn_map} (b)].
Since the phase boundaries are smoothly connected,
topological phases enclosed by such phase boundaries 
are also smoothly connected.

Let us focus on a line of $b=1$ in the phase diagram at 
$\gamma_1=\gamma_2=0$ shown in panel (b).
On this line, which we call $\eta_0$,
the correspondence between the bulk winding number $w$
defined in $g_{bulk}$
and
the existence/absence of a zero-energy edge state
under $g_{edge}$
is well established;
$w=1$ corresponds to the existence
and $w=0$ to the absence of an edge state in $g_{edge}$.
\cite{ryu}

Let us consider a more generic path $\eta$
on the phase diagram shown in panel (a).
Thanks to the smooth deformation described in the last
paragraph,
on any path $\eta$ smoothly connected to $\eta_0$,
the correspondence in the behavior of the system 
under $g_{bulk}$ and the one under $g_{edge}$
is guaranteed.
In Fig. 2(a)
the central $w=1$ region is in contact with
the right $w=0$ region by a single point $P$
located at
$(t_1,\log b)\simeq (1.44, -0.82) \equiv (t_{1c}, \log b_c)$.
For the path $\eta$ 
to be smoothly connected to $\eta_0$,
it must stay in the $w=1$ region until $t_1=t_{1c}$,
then the path must go through the point $P$
to enter directly the right $w=0$ region.
A similar argument applies on the $t_1<0$ side.
Any of such a path $\eta$
is smoothly connected to $\eta_0$,
and the corresponding bulk geometry $g_{bulk}$
is eligible for establishing the bulk-edge correspondence.

The bulk-edge correspondence
in the hermitian limit is 
based on the smooth deformation
of a loop $\rho=R_-(\beta=e^{ik})$ ($k\in[0,2\pi]$)
in the space of model parameters $\tau$.\cite{ryu}
Here, in the non-hermitian case
the same deformation must be done in the generalized parameter space
$\tilde{\tau}=\{\tau,b\}$.
In this sense
the bulk-edge correspondence
for non-hermitian topological systems
is a generalized one.

The generalized bulk-edge correspondence just proven 
gives also a clear interpretation to the pbc spectrum
shown in Fig. 1 
at the back of the obc spectrum for comparison;
recall that the number and locations
of the gap closing are different in the two spectra.
In Fig. 3 (a)
the same pbc spectrum is plotted in a broader range of $t_1$.
In panel (b) $w(t_1,b = 1)$ is plotted.
With increasing $t_1$ from the left end,  
$w(t_1,b = 1)$ changes as 
$0\rightarrow 1/2\rightarrow 1\rightarrow 1/2\rightarrow 0$.
This corresponds to the $b=1$ line in Fig. 2 (a).
Each time $w(t_1,b)$ changes, the pair of winding numbers $(w_+,w_-)$  
changes, 
and hence the system undergoes a topological phase transition. 
A  gap closing must occur at corresponding values of $t_1$. 
In panel (a), 
the central region around $t_1=0$
bounded by two gap closings at $t_1\simeq 0.3$ and at  $t_1\simeq -0.4$
falls on  the $w = 1$ region. 
This is actually the part, which is smoothly 
connected to the TI phase under obc with a pair of zero-energy edge  
state. 
The two far ends of the spectrum ($t_1\lesssim -1.8, t_1 \gtrsim 1.9$)  
correspond to the OI phase.
The remaining intermediate region 
between the inner and outer gap closings falls on the new topological phase with $w = 1/2$.
Such a phase corresponding to $(w_+, w_-)=(1,0)$ or $(0,-1)$
has no analogue in the hermitian limit, 
realizing a new topologically distinct phase
which is truly non-hermitian.
Since
these $w=1/2$ regions 
cannot be smoothly connected to
a known topological phase in the hermitian limit (characterized by edge states under obc),
we cannot characterize them in such a conventional way.
We leave further analysis on the nature of these new topological phases
to future study.

Strictly speaking, 
the proof given here is applicable to 
topological phases 
in the perturbative non-hermitian regime;
to those connected smoothly to the corresponding hermitian topological phase.
A more general proof of the bulk-edge correspondence
valid also in the non-perturbative non-hermitian regime will be given elsewhere.
\footnote{K.-I. Imura, Y. Takane, in preparation.}

Unlike our recipe employing mpbc for $g_{bulk}$
and obc for $g_{edge}$ to recover the bulk-edge correspondence,
the authors of Refs. \onlinecite{YaoWang,non-Bloch}
developed an alternative approach employing only $g_{edge}$ specified by obc. 
\cite{YaoWang,non-Bloch}
Under mpbc
the eigenstates of the system takes 
the generalized Bloch form
(\ref{beta>}).
Under obc the eigenstate 
is no longer in this form;
instead it becomes a linear combination of the Bloch form (\ref{beta>})
with different $\beta$.\cite{non-Bloch}
Ref. \onlinecite{non-Bloch} gives a recipe to find the bulk solutions compatible with obc
in the limit of $L\rightarrow\infty$,
where
$|\beta|$ is not a constant but a function of $k$.
The Brillouin zone, i.e.,
the trajectory $C_\beta$ of $\beta$,
is no longer circular
and has cusps in certain cases.
\cite{YaoWang,non-Bloch}

As for the bulk-edge correspondence,
a winding number formally similar to the ones 
in Eq. (\ref{wn_pm})
is employed
to characterize
the mapping from $C_\beta$ to a loop of $\rho=R_\pm (\beta)$. 
\cite{YaoWang,non-Bloch}
There, the integral over $k\in[0,2\pi]$ is replaced with a contour integral
along $C_\beta$.
In the derivation of these winding numbers,
left and right eigenstates of the Bloch Hamiltonian $H(\beta)$ is needed.
However, such left and right eigenstates 
are specified by a single $\beta$, so that they do not 
satisfy obc, 
while
the values of $\beta$ on the contour $C_\beta$
result from obc.
Recall that under obc
the eigenstates are not in the form of (\ref{beta>}).
In this regard, our formulation is 
more natural as every procedure is carried out based on the 
mpbc (\ref{H_mpbc}).
Furthermore, unlike the arguments of 
Refs. \onlinecite{YaoWang,non-Bloch},
we start with a system of finite size $L$ then take 
safely the limit of $L\rightarrow\infty$ at the end of the formulation, as one 
usually does in the hermitian case.

\begin{figure}
\includegraphics[width=85mm, bb=0 0 250 112]{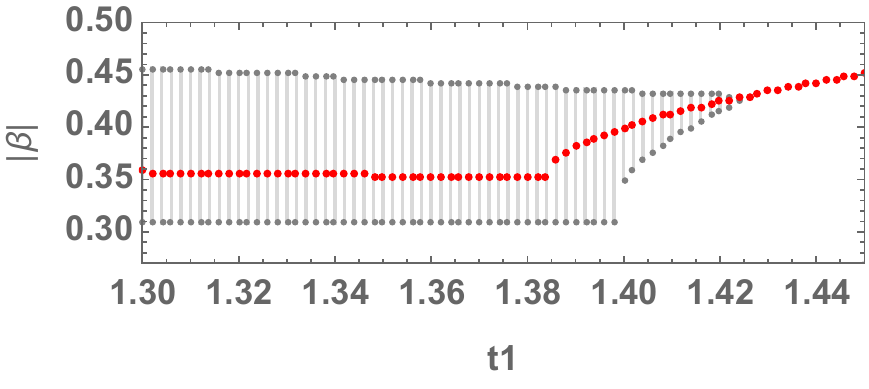}
\vspace{-2mm}
\caption{
$|\beta(E_{bot})|$
is plotted in red as a function of $t_1$
in the vicinity of
the phase transition at $t_1 = t_{1c} \simeq 1.44$.
Gray bars outline the region of $w = 1$,
which has already appeared in Fig. \ref{wn_map} (a)
though here the vertical axis is
in linear scale.
}
\label{beta_Eb}
\end{figure}

\begin{figure}
\includegraphics[width=85mm, bb=0 0 250 107]{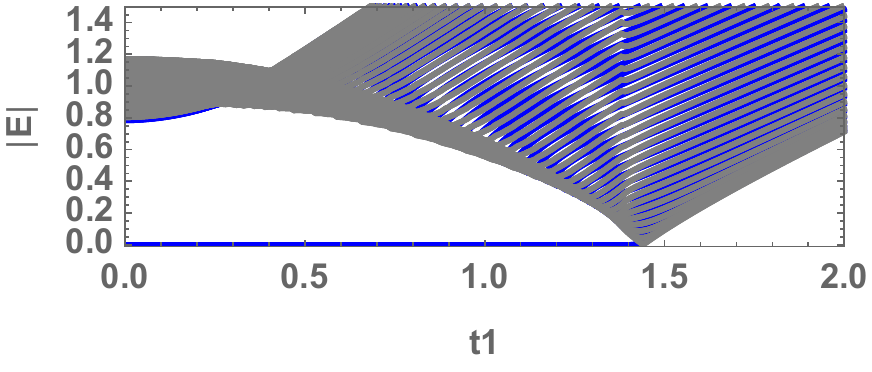}
\vspace{-2mm}
\caption{
Spectrum under mpbc.
$|E|$ is plotted in gray as a function
of $t_1$.
For comparison, the spectrum under obc (Fig. 1) is also plotted
in blue. The plot is done in a system of size $2L = 200$ for mpbc,
while $2L = 100$ for obc.
}
\label{spec_bulk2}
\end{figure}

\section{Bulk-edge correspondence on an optimal path $\bar{\eta}$}

Suppose that a path $\eta$ given on the $(t_1,b)$-plane in Fig. 2(a)
is smoothly connected to its hermitian counterpart $\eta_0$.
This guarantees the bulk-edge correspondence between $g_{bulk}$ under mpbc
with $\eta$ and $g_{edge}$ under obc as discussed in Sec IV.
In the viewpoint of spectrum, the bulk-edge correspondence only ensures
that the positions of the gap closing in $g_{bulk}$
are identical with those in $g_{edge}$.
It is not 
mandatory that the mpbc spectrum on $\eta$ reproduces
the bulk part of the obc spectrum completely.


Still it will not be useless to consider the optimal path $\bar{\eta}$,
among an infinite number of possible paths,
chosen such that the mpbc spectrum on $\bar{\eta}$
reproduces the bottom of the energy band in the bulk spectrum under obc.
Since the gap closing is a property of the band bottom,
this ensures that the positions of the gap closing are also reproduced.
Despite that $\bar{\eta}$ is not the only but a possible choice of $\eta$,
$\bar{\eta}$ has an advantage that it eases to follow the evolution
of the spectrum from the one under $g_{bulk}$ to the one under $g_{edge}$.
This is the simplest and most direct way to establish
the bulk-edge correspondence.


In determining $\bar{\eta}$, note that the trajectory $C_\beta$ of $\beta$
defines the bulk energy band of $|E|\in [E_{bot},E_{top}]$ under obc,
\cite{non-Bloch}
indicating that $\beta$ should be regarded as a function of $E$.
To reproduce the spectrum at the band bottom $E_{bot}$ under obc,
we need to tune the parameter $b$ to be consistent with
the value of $\beta$ corresponding to $E_{bot}$ at each $t_1$.
Thus, $b$ should be determined as
\begin{equation}
  \label{Eb}
  b= |\beta (E_{bot})|.
\end{equation}
As $\beta (E_{bot})$ varies as a function of $t_1$,
$b$ determined as Eq. (\ref{Eb})
defines a path $\bar{\eta}$ in the parameter space $(t_1,b)$.


The path $\bar{\eta}$ defined in this way
is smoothly connected to the one in the hermitian limit $\eta_0$.
Fig. \ref{beta_Eb} shows the value of $|\beta (E_{bot})|$ 
against the $w=1$ region in the $(t_1,b)$-space.
It shows that
within the accuracy of the numerical computation
the value of $|\beta (E_{bot})|$ is always found in the
$w=1$ region
until the very end of this region
$P$
located at $(t_1,b)= (t_{1c}, b_c)$.
After passing through the point $P$ the path $\bar{\eta}$ specified 
as $b=|\beta (E_{bot})|$
gets directly into the $w=0$ region (not shown).


\begin{figure}
\includegraphics[width=85mm, bb=0 0 250 107]{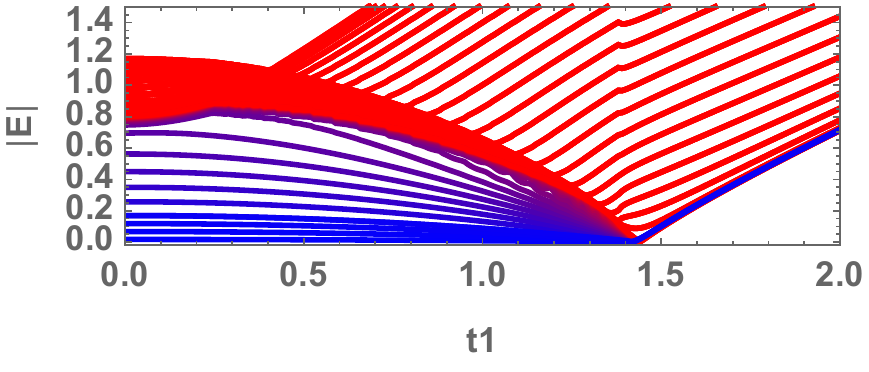}
\\
\vspace{2mm}
\includegraphics[width=85mm, bb=0 0 250 96]{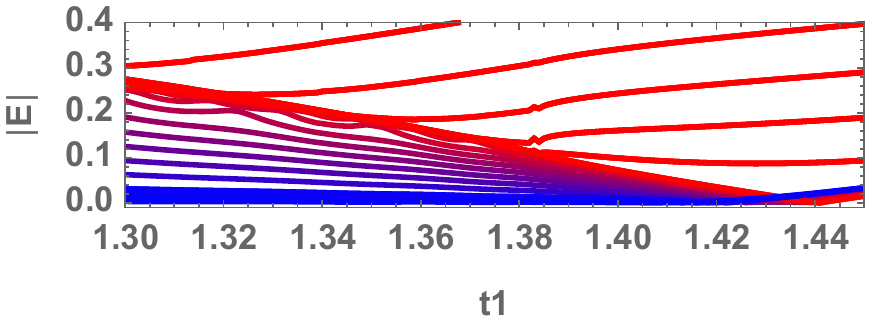}
\vspace{-2mm}
\caption{
Evolution of the spectrum from mpbc to obc. 
$|E|$ is plotted
as a function of $t_1$
at $2L = 100$. 
The mpbc (\ref{H_mpbc}) with $b=|\beta(E_{bot})|$ is
employed and then gradually switched off 
[$\delta$ in Eq. (\ref{delta}) varies from 1 to 0].
At $\delta$ away from the mpbc value ($\delta=1$)
only data from
the bottom of the spectrum is kept and added to the spectrum under mpbc
(Fig. 5) in a color that varies continuously from red (mpbc) to blue (obc).
Bottom: A detailed plot of the top panel in the vicinity of the phase
transition ($t_1 = t_{1c}$). 
}
\label{pbc2obc}
\end{figure}

Fig. \ref{spec_bulk2} shows the spectrum under mpbc with $b=|\beta (E_{bot})|$.
At sight
it looks reproducing the bulk part of the obc spectrum fairly well.
Strictly speaking,
the value of $|\beta (E_{bot})|$ is justified only in the $L\rightarrow\infty$ limit,
while the obc spectrum is calculated at a finite $L$
($2L=100$),
so that this agreement is approximate.
Also if one examines the spectrum more closely,
one can recognize that the agreement is limited to the vicinity of the band bottom.
But still, 
such an agreement is
advantageous for making the bulk-edge correspondence 
intuitively
accessible.

Assuming that the mpbc (\ref{H_mpbc}) with $b$ as given in Eq. (\ref{Eb}) is initially imposed,
let us gradually switch it off:
\begin{equation}
H_\delta=H_{obc}+
\delta \left[H_{\rm NN}^{mpbc}+H_{\rm 3NN}^{mpbc}\right]_{b=|\beta(E_{bot})|}
\label{delta}
\end{equation}
with $\delta$ varied from 1 to 0.
In Fig. \ref{pbc2obc}
one can clearly see that
a branch of spectrum detached from
the bottom of the energy band 
in the purely bulk mpbc spectrum (situation of Fig. \ref{spec_bulk2} at $\delta=1$)
evolves continuously into the zero-energy edge state at obc ($\delta=0$).
\footnote{A wrinkle-like feature around $t_1\simeq 1.38$ is due to the disappearance of cusps
in the generalized Brillouin zone
\cite{non-Bloch}
}
The smooth evolution is highlighted by a continuous change
of plot colors used for that branch.
This intuitive form of bulk-edge correspondence 
is first made possible
thanks to mpbc.

\section{Concluding remarks}

The non-hermitian skin effect is specific to non-hermitian systems
under open boundary conditions (obc).
Under obc
the wave function exhibits a typical exponential dependence
and is localized near the boundary of the system.
This non-hermitian skin effect is influential 
even to the topological nature of the system.
The gap closing specifying
the phase boundary between TI and OI phases under obc
becomes unrecognizable in the spectrum 
under ordinary periodic boundary condition (pbc).
The apparent failure of the bulk-edge correspondence in the ordinary pbc
stems from the fact that it is incompatible with the skin effect. 

Here, we have proposed 
a modified periodic boundary condition (mpbc),
which takes a proper account of 
this effect in a closed geometry.
The mpbc 
maintains the topological features 
that arise under obc, while
the remnant of the skin effect can be 
clearly seen in the eigenfunctions under mpbc
in spite of the closed geometry.

In the mpbc
the non-hermitian skin effect is taken into account
through a parameter $b$
that specifies the boundary condition.
The correspondence between
bulk and edge has been established in a generalized parameter space 
$\tilde{\tau}=\{\tau,b\}$,
where
$\tau$ is the space of model parameters.
In this sense 
the concept of bulk-edge correspondence is generalized in non-hermitian systems.
Following the arguments of Ref. \onlinecite{ryu},
we have proven this generalized bulk-edge correspondence,
which constitutes the main result of the paper.
The mpbc
also enables us to conceive the idea of bulk-edge correspondence
in a generic non-hermitian system
intuitively
as an evolution of the spectrum
in the course of the continuous change of the boundary condition
from mpbc to obc.

\acknowledgements
The authors thank 
Kohei Kawabata, Naomichi Hatano and Hideaki Obuse
for many useful discussions and correspondences.
This work has been supported by
JSPS KAKENHI Grant No. 15K05131, 18H03683, 15H03700, and 18K03460.

\bibliography{drd4bib_r2}

\end{document}